\documentstyle[11pt,newpasp,twoside,epsf]{article}
\def\cm{{\rm\thinspace cm}}

\def\erg{{\rm\thinspace erg}}

\def\s{{\rm\thinspace s}}
\def\erg{{\rm\thinspace erg}}

\def\ergpcmsqps{\hbox{$\erg\cm^{-2}\s^{-1}\,$}}

\def\ergps{\hbox{$\erg\s^{-1}\,$}}

\def\pcmsq{\hbox{$\cm^{-2}\,$}}

\def\approxlt{\mathrel{\spose{\lower 3pt\hbox{$\sim$}}
        \raise 2.0pt\hbox{$<$}}}
\def\approxgt{\mathrel{\spose{\lower 3pt\hbox{$\sim$}}
        \raise 2.0pt\hbox{$>$}}}

\def\grs1915{GRS~1915+105}

\def\edcomment#1{\iffalse\marginpar{\raggedright\sl#1\/}\else\relax\fi}
\reversemarginpar

\begin{document}
\title{Broad P Cygni X-ray Lines from Circinus X-1}
 \author{Norbert S. Schulz}
\affil{Massachusetts Institute of Technology, Center for Space Research, Cambridge, MA, 02139}
\author{William N. Brandt}
\affil{The Pennsylvania State University, Department of Astronomy and Astrophysics, University Park, PA, 16802}

\begin{abstract}
Chandra High Energy Transmission Grating Spectrometer (HETGS) observations of the
extremely luminous and violent X-ray binary Cir X-1 reveal many X-ray emission lines
showing broad P Cygni profiles.
These are from H-like and He-like ions of Ne, Mg, Si, S, Ar, and Fe. The lines
originate in a high-velocity (up to 2000 km s$^{-1}$) outflow, which may be an
accretion disk wind.
Cir X-1 has a radio jet, and during some periastron passages of the
neutron star (orbital phase = 0.0), it flares at radio, infrared and X-ray
wavelengths. The X-ray P Cygni profiles were observed during this phase,
where we also expect the strongest mass transfer from the companion star.
We present new results on the time variability of the P Cygni 
profiles and discuss their general importance. 
\end{abstract}

\section{Introduction}

The X-ray binary Circinus X-1 contains a neutron star orbiting a companion star every 16.5 days.
The X-ray emission from compact binaries depends strongly on the accretion flow 
structure and the nature of the compact star. In the case of \hbox{Cir X-1} it is well established 
that the compact star is a neutron star, since EXOSAT observations revealed type I X-ray bursts
from this system (Tennant et al. 1986). These bursts are the result of explosive thermonuclear burning
on the surface of a relatively weakly magnetized ($< 10^{10}$ Gauss) neutron star. The nature of its stellar
companion, however, is not as well established because \hbox{Cir X-1} lies in a region where a large amount 
of dust and absorbing material blocks the line of sight. 
\hbox{Cir X-1} is probably a low mass X-ray binary (LMXB), which is based 
on the identification with a rather faint optical counterpart (Moneti 1992) and more strongly
on X-ray spectral variations in the color-color diagram observed with RXTE (Shirey et
al. 1999a). 

The binary orbit is thought to be quite eccentric (e.g. Tauris et al. 1999), 
and the neutron star accretes from its stellar companion via Roche lobe overflow 
only near periastron passages (``zero phase''). This explains some of the
rapid variability and the up to super-Eddington X-ray luminosities \hbox{Cir X-1} exhibits on a quasi-regular
basis (e.g. Shirey et al. 1999b and references therein). The first observations that offered moderately high
spectral resolution in the X-ray band were performed with ASCA (Brandt et al. 1996).  
The ASCA spectral analysis showed that during some flux transitions the luminosity of the system 
does not change strongly and the variations could be explained by partial-covering absorption. The data also
indicate that the accretion disk is viewed in a rather edge-on manner. 
Recently radio jets were observed in this system (Stewart et al. 1993; Fender et al. 1999), which so far
have only been observed in the so-called microquasars.

In this presentation we show the first high-resolution X-ray spectra of \hbox{Cir X-1}    
obtained by Chandra. A more in depth analysis of these is presented by
Brandt $\&$ Schulz (2000) and Schulz $\&$ Brandt (2001).    

\section{Observations}

We observed Cir X-1 with the Chandra High Energy Transmission Grating Spectrometer
(HETGS) on 2000 February 29 at 22:09:50 UT for 32 ks. This time corresponds to
zero phase using the ephemeris in Glass (1994). The HETGS allows one to observe simultanously at two
different spectral resolutions:  the Medium Energy Gratings (MEG) with a resolution of 0.02 ~\AA\
are optimized for a bandpass of 2.5--31~\AA\, (0.4--5~keV) and the High Energy Gratings (HEG) 
with a resolution of 0.01 ~\AA\ are optimized for a bandpass of 1.5--13~\AA\, (0.9--8~keV). 
In addition, the resolution for higher orders increases by a factor $n$ for the $n$th order.

The source appeared very bright, and we registered a total of
1.4$\times 10^6$ events in MEG $-1$st order, 3.5$\times 10^5$ events in MEG $-3$rd order, and
1.5$\times 10^6$ events in HEG +1st order. Due to a misplacement of a subarray window during
the observation, we only have  negative MEG and positive HEG orders. 
The source was bright enough to suffer from severe photon pileup in the first-order spectra and
we therefore exclude the bandpass below 10 ~\AA\ in the MEG 1st order, above 2.5 ~\AA\ and below
6 ~\AA\ in HEG 1st order, and below 2 ~\AA\ in the MEG 3rd order. The last exclusion is due to the
effect that MEG 1st order pileup events contaminate the 3rd order spectrum.

A most critical item is the wavelength scale calibration. Here we determined the 0th order
position to an accuracy of 0.3 pixels ($< 0.004$~\AA\ in MEG $-1$st, 0.002 ~\AA\ in HEG +1st orders)
and also correct for the actual detector pixel size, which is currently slightly overvalued
in the standard processing pipelines. The total wavelength scale error is then well below
0.1$\%$. The effective areas are currently known to about 10$\%$ in the 1st orders and about 15$\%$ in the 
3rd orders.
 
\begin{figure*}[t]
\plotone{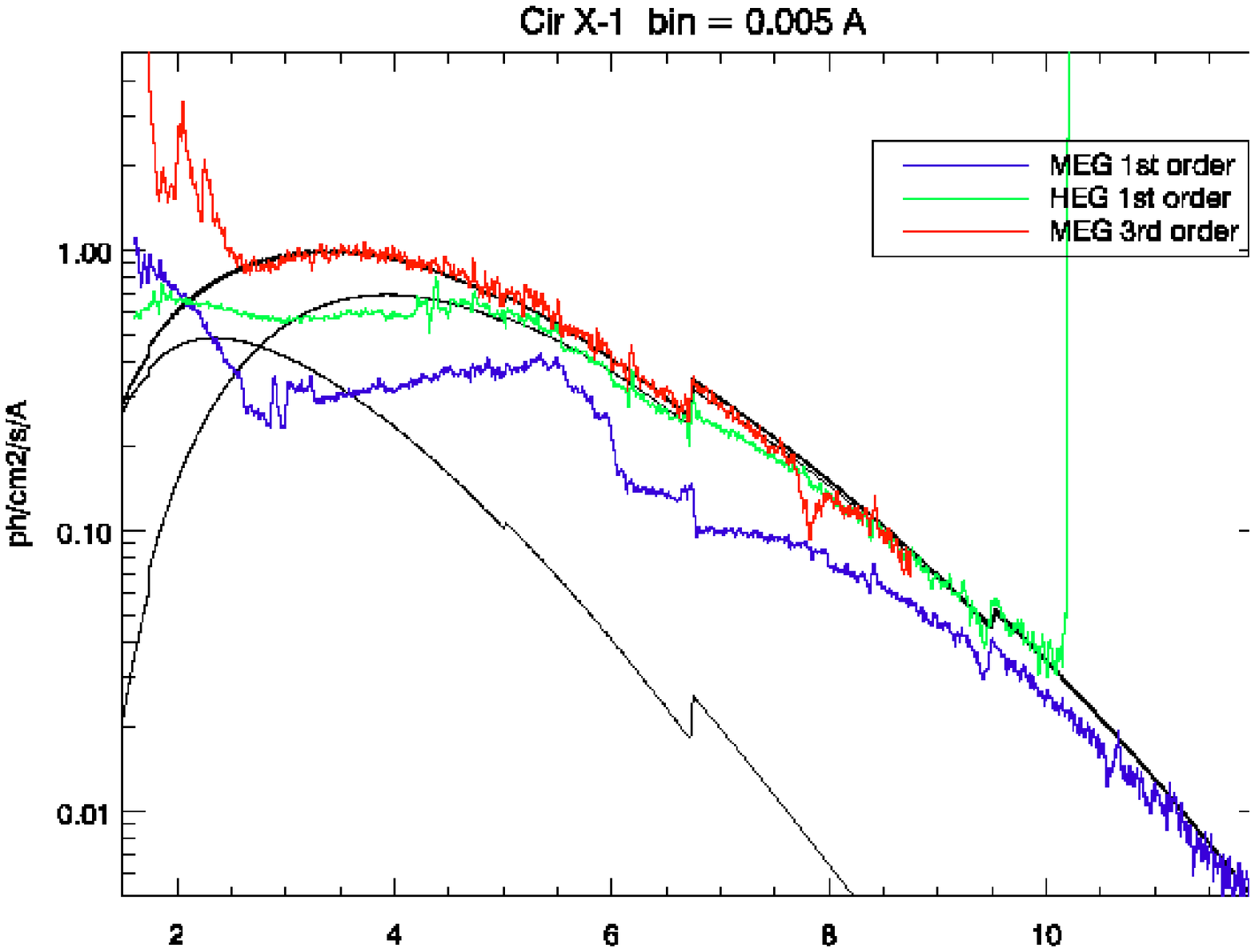} 
\vspace{-0.4in}
\centerline{\it wavelength (\AA) \rm}
\caption{Effective area and exposure corrected spectra from MEG $-1$st order (bottom),
HEG + 1st order (middle), and MEG $-3$rd order (top). Most of the residual difference
is caused by photon pileup.}
\label{cirx1-fit}
\end{figure*}

\section{The Spectral Continuum}

The analysis of the spectral continuum is complicated due to the fact that for now we have
to recognize regions affected by pileup as well as take into account some contamination from
0th order scattering. In order to simplify the analysis we perform the spectral fits after we correct
the spectra for exposure and instrumental effects. This is quite legitimate in the case of
the HETGS, because here the instrumental response is comparable to the pixel size of the detector.
We also restrict the analysis to a single model type, where we apply a double
blackbody spectrum with partial-covering absorption as was successfully done
in the analysis of ASCA data by Brandt et al. (1996). 

Figure 1 shows the reduced HETGS spectra for the MEG $-1$st order (bottom, black),
the HEG +1st order (middle, gray), and the MEG $-3$rd order (top, black). Because of pileup
these spectra do not agree with each other thoughout the entire bandpass. We only use
the MEG 3rd order spectrum above 2.3 ~\AA\ for the fit and use the HEG and MEG 1st order
spectra to further constrain the fit above 8 ~\AA\ and 11 ~\AA\, respectively. The results are summarized
in Table 1. The two blackbodies as well as their sum are also displayed in Figure 1. 

\begin{table}
\begin{center}
\begin{tabular}{lccc}
\multicolumn{4}{c}{\sc Table 1: Double blackbody spectral fit results} \\
\hline
\hline
\multicolumn{4}{c}{\sc Interstellar Column density: (2.0$\pm 0.1) \times 10 ^{22} \pcmsq$} \\
Model& N$_H$ (10$^{22} \pcmsq$) & Cov. fraction & $kT$ (keV) \\ 
\hline
Blackbody 1 (part. cov.) & 82$\pm$4 & 0.45 & 0.66$\pm$0.17 \\
Blackbody 2 (part. cov.) & 2.3$\pm$0.2 & 0.89 & 1.22$\pm$0.28 \\
\hline
\end{tabular}
\end{center}
\end{table}

The spectral fit parameters are quite comparable to the ASCA results in the high count rate state
(Brandt et al. 1996). With a resulting X-ray flux of 1.8$\times 10^{-8}$ \ergpcmsqps and
a distance of 6 kpc, we determine a 2--8 keV luminosity of 9.7$\times 10^{37}$ \ergps.

\begin{figure*}[t]
\plotone{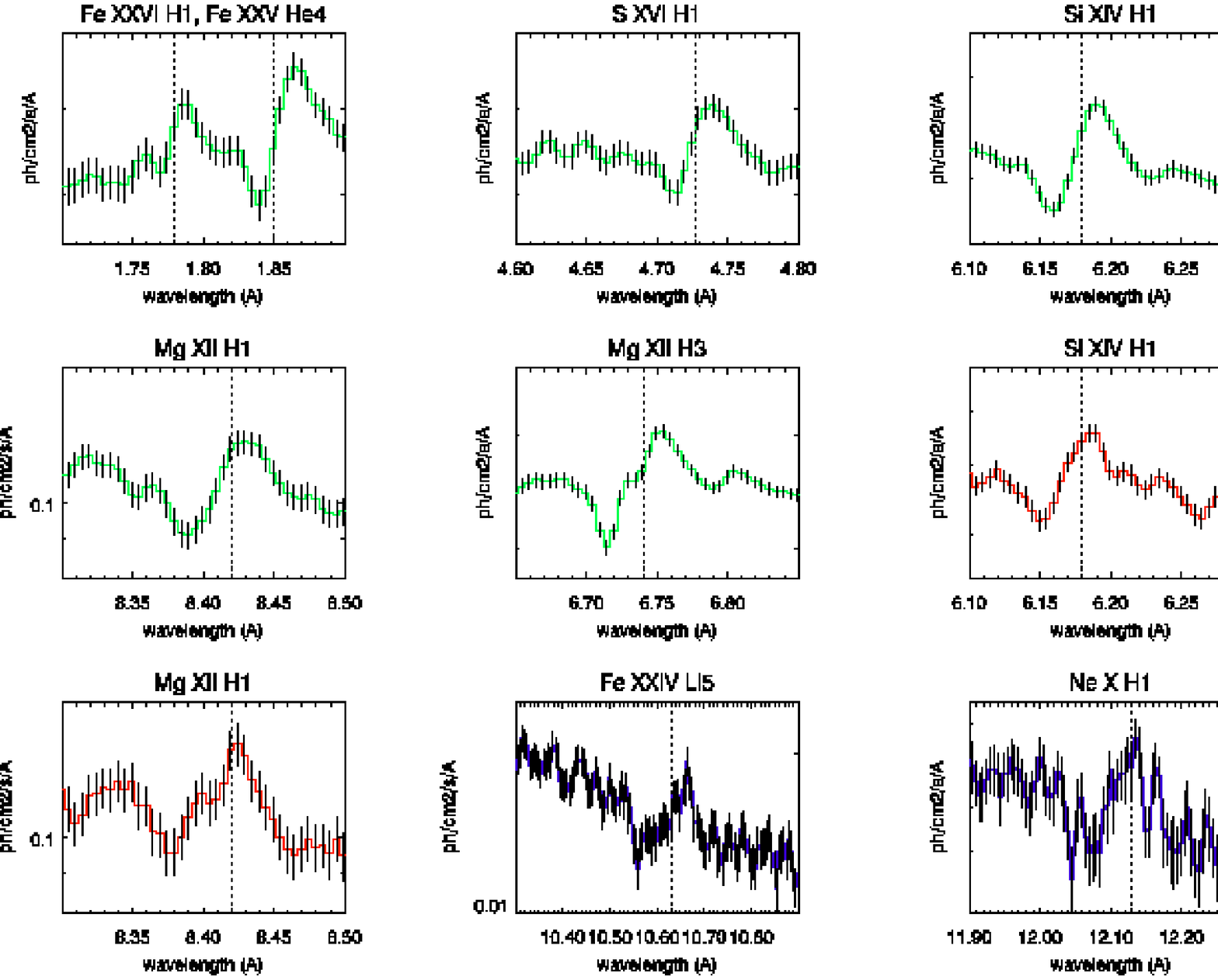}
\vspace{0.00001in}
\caption{A selection of P Cygni line profiles for various inner-shell transitions, reading from left to
right and top to bottom row:  
profiles from HEG 1st order include Fe XXVI and XXV (He-like), S XVI, Si XIV, 
Mg XII , Mg XII (Lyman $\gamma$); profiles from MEG 3rd order include Si XIV and Mg XII; profiles
from MEG 1st order include Fe XXIV (Li-like) and Ne X ions. If not mentioned otherwise,
all lines are Lyman $\alpha$ transitions.}
\label{cirx1-profiles}
\end{figure*}

\section{P Cygni Lines}

The spectra also show significant emission lines from hydrogen (H-like) and helium (He-like)
like ions. Thus far we have identified lines from Ne, Mg, Si, S, and Fe. The lines are broad
and in addition show blue shifted absorption components, which are quite typical for P Cygni
line profiles. Figure  2  shows various examples of these lines from Fe XXV and XXVI (top, left)
to {Ne X} (bottom, right). The profiles on average appear quite symmetric and the line widths in 
absorption as well as emission are comparable. We determined the wavelength difference between
the peak of the emission and the valley of the absorption to be within 0.02 and 0.10 ~\AA\
depending on the ion. Once we remove the contribution of the instrumental response and 
calculate a Doppler velocity, we get
total (red plus blue shifted) velocities between 800 and 3700 km s$^{-1}$.

We also have performed some analysis of variations of the profiles with time. Figure 3 shows a time
sequence of the Si XIV Lyman $\alpha$ line, where we integrated each spectrum for 5000 s and 
shifted the start time by 1000 s each time (this is necessary in order to have enough exposure for
each spectral segment). Although these spectral segments are not entirely independent
of each other, we observe changes in the profile shape on time scales of one to a few hours.
In the case of the Si XIV line, for example, the absorption vanishes in the middle of the observation almost
entirely and emission dominates. The other profiles show similar time variability, and it is 
not yet clear from our observation how much of this variability is correlated with the total
flux of the system.

\vspace{-0.1in}

\section{Summary}

The discovery of P Cygni X-ray line profiles from Cir X-1 fits well into predictions of previous
observations in that it is likely we view the system in a relatively edge-on manner. In this view it is
not unreasonable to interpret these profiles in terms of a high-velocity equatorial 
outflow. 
In the case of Cir X-1 this wind is launched in a layer of moderate temperature ($> 10^6$ K)
where atomic heating and cooling processes dominate over Compton processes. The line emission we
observe is quite similar to what Raymond (1993) predicted for such a configuration. 
Material from this layer could be thermally or radiatively driven outwards in a wind,
and the P Cygni profiles may represent this wind (e.g. Begelman, McKee, $\&$ Shields 1983).
In the case of Cir X-1 such an outflow must be quite dense to keep the ionization
parameter small enough so the ions can survive (Kallman $\&$ McCray 1982). This is plausible
for densities larger than a few 10$^{15}$ cm$^{-3}$ (Brandt $\&$ Schulz 2000). 
  
The first observation of P Cygni profiles in high-resolution X-ray spectra also has implications
for high-energy astrophysics in general, since it demonstrates that with Chandra we are
now able to study the dynamics of systems as was previously possible only at longer wavelengths.
 Clearly we should observe {P Cygni} profiles from other systems with outflows, especially
from microquasars or systems involving a stellar wind. In fact, following the detection of such
profiles in Cir X-1, Liedahl et al. (2000) recently reported on such profiles in HETGS spectra
of Cyg X-3 during outburst. Here the profiles are interpreted as the interaction of a high-velocity
wind from a massive Wolf-Rayet star with an accreting neutron star. In the case of Cir X-1
such an origin appears unlikely, because the probable low-mass companion should not exhibit such 
a strong and high-velocity wind. Also, similar line profiles were
discovered in UV spectra of cataclysmic variables possessing an accretion disk (e.g., C\'ordova $\&$
Howarth 1987). However, these lines are modelled in terms of moderately collimated
polar outflows (e.g., Knigge $\&$ Drew 1997) and thus are probably not related to the
phenomenon observed in Cir X-1.

\begin{figure*}[b]
\vspace{1.6in}
\plotfiddle{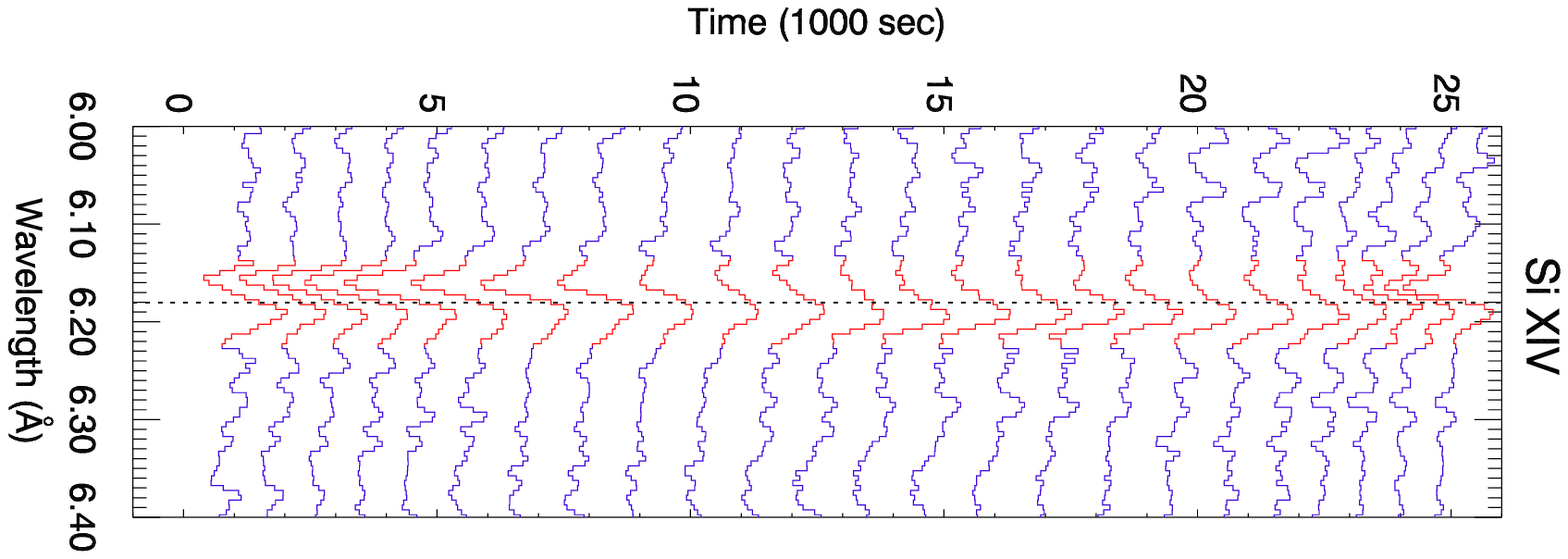}{0.1cm}{90}{70}{70}{400}{-55}
\vspace{0.00001in}
\caption{Time sequence of the Si XIV Lyman $\alpha$ P Cygni line.}
\label{cirx1-si}
\end{figure*}

\end{document}